\documentclass[a4paper]{article}

\usepackage{authblk}
\usepackage[T1]{fontenc}    
\usepackage{hyperref}       
\usepackage{url}            
\usepackage{amsfonts}       
\usepackage{amsmath}
\usepackage{microtype}      
\usepackage{graphicx}
\usepackage{doi}
\usepackage{cleveref}
\usepackage[style=numeric, sorting=none]{biblatex}
\usepackage[version=3]{mhchem}
\usepackage{siunitx}
\DeclareSIUnit\oersted{Oe}
\DeclareSIUnit\phinull{\Phi_0}
\DeclareSIUnit\mub{\mu_B}
\DeclareSIUnit\micro{\mu}

\usepackage{subcaption}
\newcommand{\labelphantom}[1]{%
  \parbox{0pt}{\phantomsubcaption\label{#1}}%
}

\title{Large output voltage to magnetic flux change in nanoSQUIDs based on direct-write Focused Ion Beam Induced Deposition technique}

\author[a]{Fabian Sigloch}
\author[a,b]{Soraya Sangiao}
\author[a]{Pablo Orús}
\author[a,b,*]{José María de Teresa}

\affil[a]{Instituto de Nanociencia y Materiales de Aragon (INMA), Universidad de Zaragoza-CSIC, 50009 Zaragoza, Spain}
\affil[b]{Laboratorio de Microscop{\'i}as Avanzadas (LMA), Universidad de Zaragoza, 50018, Spain}
\affil[*]{E-Mail: \href{deteresa@unizar.es}{deteresa@unizar.es}}

\hypersetup{
    pdftitle={Large output voltage to magnetic flux change in naoSQUIDs based on direct-write Focused Ion Beam Induced Deposition technique},
    pdfauthor={Fabian Sigloch, Soraya Sangiao, Pablo Orus, Jose Maria de Teresa},
}

\begin{document}
\maketitle

\begin{abstract}
    \noindent NanoSQUIDs are quantum sensors that excel in detecting a small change in magnetic flux with high sensitivity and high spatial resolution. 
    Here, we employ resist-free direct-write \ce{Ga+} Focused Ion Beam Induced Deposition (FIBID) techniques to grow W-C nanoSQUIDs, and we investigate their electrical response to changes in the magnetic flux. 
Remarkably, FIBID allows the fast (\SI{3}{\minute}) growth of $\SI{700}{\nano\metre}\times\SI{300}{\nano\metre}$ Dayem-bridge nanoSQUIDs based on narrow nanowires (\SI{50}{\nano\metre} wide) that act as Josephson junctions. 
    The observed transfer coefficient (output voltage to magnetic flux change) is very high (up to \SI{1301}{\micro\volt\per\phinull}), which correlates with the high resistivity of W-C in the normal state. 
    We discuss here the potential of this approach to reduce the active area of the nanoSQUIDs to gain spatial resolution as well as their integration on cantilevers for scanning-SQUID applications.
\end{abstract}

\section{Introduction}
\label{sec:Introduction}

Direct current- (dc-) Superconducting Quantum Interference Devices (SQUIDs) are magnetic flux sensors that attain an unrivaled sensitivity \cite{Koelle2017, Clarke2004} by exploiting the physical effects of magnetic flux quantization \cite{London1950} and Josephson effect \cite{Josephson1962}. 
A dc-SQUID consists of a superconducting ring interrupted by two Josephson junctions (JJs), one on either side (\cref{sfig:Introduction-SQUID}).
A JJ is formed by a superconductor (S) interrupted by either an insulator (I), a normal metal (N) or a weaker superconductor (s) resulting, respectively, in a SIS-, SNS- or SsS-junction that is capable of carrying a superconducting Josephson current, $I_\text{J}$.
The flux quantization within the superconducting ring is attained by the induction of a loop current, $J$, which either opposes or supports the external magnetic field, lowering or rising the flux threading the ring to an integer multiple of the magnetic flux quantum, $\Phi_0$.
The bias current, $I_b$, injected into one of the arms of the SQUID splits into two phase-sensitive Josephson currents running simultaneously through both of the JJs and interfering in the second arm.

\begin{figure}[!hp]
    \centering
    \includegraphics[width=.5\linewidth]{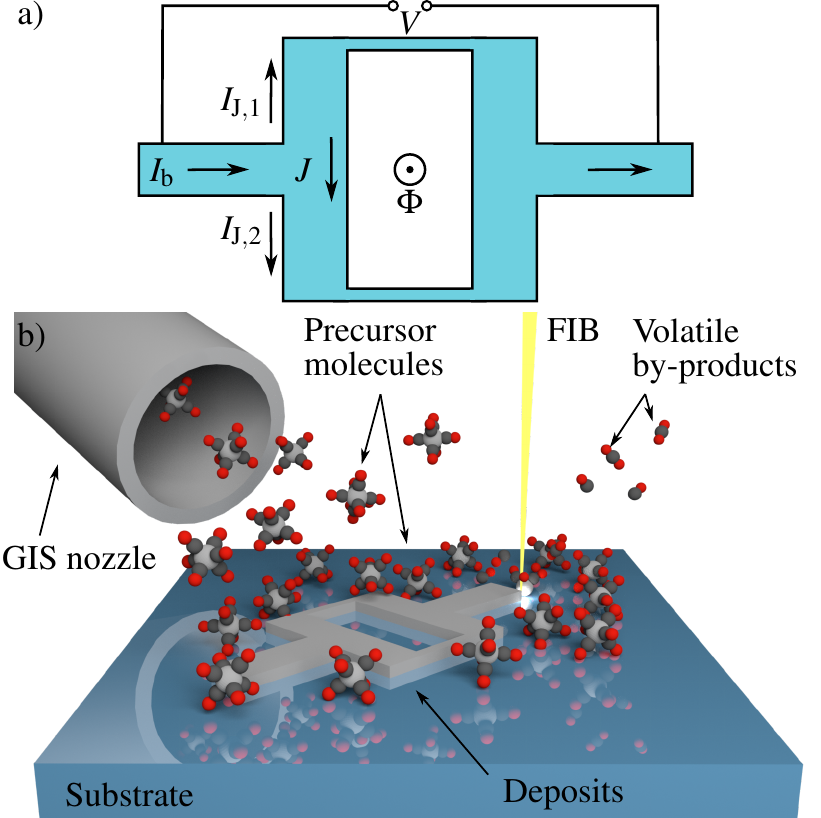}
    \labelphantom{sfig:Introduction-SQUID}
    \labelphantom{sfig:Introduction-FIBID}

    \caption{%
        a) Schematic working principle of a dc-SQUID. %
        A constant bias current $I_\text{b}$ is injected. %
        It splits up into two Josephson currents $I_{\text{J},i}$ running through the Josephson junctions on either side of the SQUID. %
        A loop current $J$ is induced to lower or raise the flux $\Phi$ threading the SQUID loop to an integer multiple of the magnetic flux quantum $\Phi_0$. %
        For $I_\text{b}\sim I_\text{c}$ a flux-dependent voltage $V$ drops across the SQUID. %
        b) Schematic working principle of Focused Ion Beam Induced Deposition (FIBID). %
        The precursor molecules are locally decomposed by irradiation with the Focused Ion Beam (FIB). %
        The metallic constituents remain on the substrate constituting the deposit, the volatile by-products are removed by the vacuum system of the instrument. %
    }
    \label{fig:Introduction}
\end{figure}

The sensitivity of SQUIDs is limited by the flux noise, $S_\Phi$, and the SQUID inductance, $L$.
The flux noise can be reduced by downsizing of the dimensions of the JJs \cite{Voss1980}, whereas the inductance can be reduced by decreasing the size of the effective inner SQUID area \cite{Ketchen1984}.
Ever since this revelation a lot of effort has been brought forth and various methods have been developed to fabricate \emph{nanoSQUIDs} with ever lower geometrical dimensions and thus higher sensitivity \cite{Wernsdorfer1995, Gallop2002, Lam2003}.

While SQUIDs with SIS- or SNS-junctions commonly require a sandwich-type structure composed of multiple layers, SQUIDs based on SsS-junctions are planar and fabricated in a single layer.
In this approach the JJs are realized by two constrictions in the SQUID loop, the Dayem Bridges (DBs) \cite{Dayem1964}, forming regions with lower $I_\text{c}$ in the superconducting material.
The low thickness of the planar DB-nanoSQUIDs makes them insusceptible to in-plane fields and enables for good coupling to magnetic nanoparticles.
However, the dissipation of heat in the normal-conducting state of the DBs yields a hysteretic I-V characteristic in materials with high $I_\text{c}$.
Generally, the kinetic inductance, $L_\text{kin}$ is high and can dominate the total inductance of the SQUID \cite{Koelle2017}.

Conventionally, the design for a nanoSQUID is transferred to a resist by means of Electron Beam Lithography (EBL).
The structure can either be deposited via evaporation of a superconducting material followed by lift-off or etched from a previously patterned superconducting film \cite{Wernsdorfer2009}.
Processes based on EBL are well established and allow for complicated geometries with linewidths down to \SI{30}{\nano\metre} \cite{Voss1980}. 
NanoSQUIDs based on DBs have been fabricated with an inner loop area of down to $\SI{200}{\nano\metre}\times\SI{100}{\nano\metre}$ and a spin sensitivity of $\sim\SI{10000}{\mub\per\sqrt{\hertz}}$ by means of EBL \cite{Wernsdorfer2009}.
However, the resist-based EBL process requires multiple fabrication steps making it a time-consuming approach.
A homogeneous film of the resist is obtained by spin-coating, requiring a large, flat substrate.
Furthermore, the resulting structures are not perfectly symmetric and suffer from irregular edges \cite{Foley2009}.

Novel, sophisticated fabrication methods such as variable thickness DBs \cite{Bouchiat2001}, superconducting \ce{Nb}/\ce{Al} bilayers \cite{Hazra2014} and normal-conducting heat-sinks \cite{Lam2003} could further increase the sensitivity, but also add to the complexity of the fabrication process.
The currently smallest, most sensitive nanoSQUIDs are based on a complicated process of directional evaporation of a superconducting material onto a pulled quartz tube.
Vasyukov \emph{et al.} fabricated a circular SQUID with a diameter of \SI{50}{\nano\metre}, resulting in an inductance of \SI{5.8}{\pico\henry} and a spin sensitivity below \SI{1}{\mub\per\sqrt{\hertz}} making it capable of the detection of the spin of a single electron \cite{Vasyukov2013}.

A different approach to creating superconducting devices using direct-write techniques is to start from a superconducting thin film and perform a FIB irradiation process to locally modify the electronic properties. 
This approach has allowed, for example, for the creation of high quality Josephson superconducting tunnelling junctions by irradiation with a focused \ce{He+} ion beam \cite{Cybart2015}.
NanoSQUIDs with a DB-width of \SI{30}{\nano\metre} and a loop size of \SI{1}{\micro\metre} were fabricated by \ce{Ga+} FIB milling of a previously patterned \ce{Nb} film in 1980 for the first time \cite{Voss1980}.
Recently, M. Wyss \emph{et al.} used this technique to fabricate a SQUID on the tip of a capped AFM cantilever with a field sensitivity of \SI{9.5}{\nano\tesla\per\sqrt{\hertz}} \cite{Wyss2021}.
However, the \ce{Nb} in the DBs deteriorates due to the implantation of \ce{Ga} and amorphization that occurs in the surface and up to \SI{30}{\nano\metre} below it.

An alternative to resist-based techniques or directional evaporation are direct-write techniques, such as Focused Electron / Ion Induced Deposition (FEBID / FIBID), which constitute versatile techniques for the fabrication of nanostructures on substrates of arbitrary size and topography \cite{Utke2008, Orus2020a}.
These techniques do not require the use of a resist and the entire nanostructure can be fabricated in a single step. 
Typically performed in either dedicated FIB instruments or in FIB/Scanning Electron Microscope (SEM) equipments, which host columns of both ions and electrons, the procedure begins by introducing a gaseous precursor containing the element of interest into the process chamber, which then adsorbs on the substrate.
Upon local irradiation of the adsorbed molecules with the focused beam, the precursor is decomposed into a non-volatile constituent, which permanently remains deposited on the surface, and into volatile by-products that are pumped away by the vacuum system of the instrument.
The resulting deposit is patterned following the shape of the scan traced by the beam (\cref{sfig:Introduction-FIBID}).
In the case of FIBID, concurrently with the deposition of the desired material, the FIB modifies the exposed material by ion implantation, amorphization and sputtering.
In absence of a precursor gas these effects can be used to locally modify the physical properties of a given sample or to locally remove material by milling \cite{Giannuzzi1999, Petit2005}.

The ability of FEBID/FIBID techniques to pattern very small features, together with their versatility for patterning on unconventional non-planar surfaces, and the well-established superconducting properties of \ce{W\hyphen C} or \ce{Nb\hyphen C} based nanodeposits created either by FEBID \cite{Blom2021} or FIBID \cite{Porrati2019}, make these techniques very promising for the fabrication of nanoSQUIDs in a single writing step. 

\begin{figure*}[!ht]
    \centering
    \includegraphics[width=\linewidth]{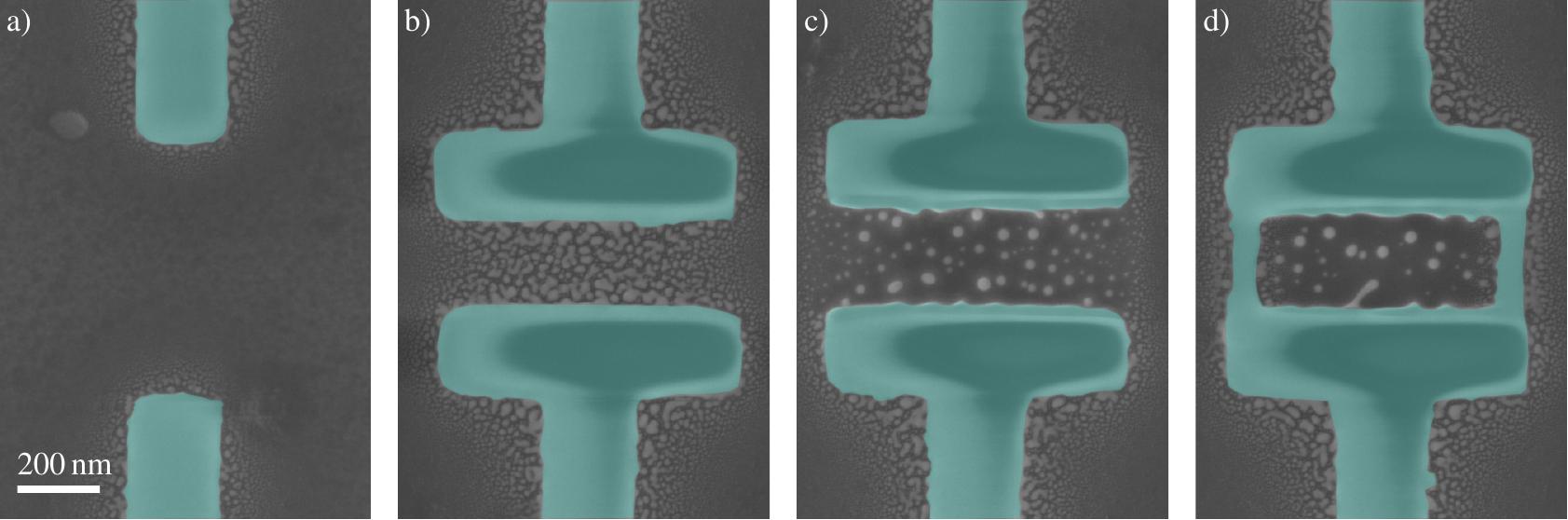}
    \labelphantom{sfig:Fabrication-a}
    \labelphantom{sfig:Fabrication-b}
    \labelphantom{sfig:Fabrication-c}
    \labelphantom{sfig:Fabrication-d}
    \caption{ %
        Artificially-colored SEM images of the fabrication procedure of a \ce{W\hyphen C}-SQUID by FIBID. %
        Blue represents the \ce{W\hyphen C} material. %
        a) First the leads connecting the SQUID to the \ce{Au} leads are deposited. %
        b) Secondly, two rectangular shapes are deposited.
        c) In the third step, the halo residing in the inner loop area is removed by a short step of FIB milling. %
        d) In the last step, narrow (\SI{50}{\nano\metre} wide) nanowires forming the Dayem bridges are grown. %
    }
    \label{fig:Fabrication}
\end{figure*}

The superconducting properties of \ce{W\hyphen C} fabricated by \ce{Ga+} FIB irradiation of the commercially available precursor gas \ce{W(CO)6} are well studied.
Planar \ce{Ga+} FIBID \ce{W\hyphen C} deposits exhibit a critical temperature of $T_\text{c}=\SIrange{4}{5}{\kelvin}$ \cite{Sadki2005, Luxmoore2007, Cordoba2013, Spoddig2007}, 
an upper critical magnetic field of $B_{\text{c}2}=\SIrange{7}{8.5}{\tesla}$ \cite{Helzel2006, Dai2013, Sun2013} 
and a critical current density of $J_\text{c}=\SIrange{0.01}{0.1}{\mega\ampere/\centi\metre\squared}$ \cite{Luxmoore2007, Cordoba2013, Spoddig2007}. 
The London penetration depth is reported to be $\lambda_\text{L}=\SI{850}{\nano\metre}$ \cite{Guillamon2009, Orus2021} and the superconducting coherence length $\xi=\SIrange{6}{9}{\nano\metre}$ \cite{Guillamon2009, Orus2021, Sadki2004, Cordoba2013}.
Nanostructures with linewidths down to \SI{50}{\nano\metre} can be patterned with high precision and reproducibility \cite{Cordoba2013}.

Several remarkable applications for superconducting nanodevices fabricated by FIBID/FEBID have been reported thus far and are worth mentioning. 
\ce{W\hyphen C} deposits have been used to induce proximity superconducting effects on other materials \cite{Wang2009} and to study the spin polarization of magnetic materials in Andreev contacts \cite{Sangiao2011}. 
Besides, narrow W-C nanowires fabricated by \ce{Ga+} FIBID have been found to sustain long-range nonlocal flow of a single row of vortices, which could be of interest for manipulation of individual vortices in quantum technologies \cite{Cordoba2019c, Golod2015}, and allow tuning the value of the critical current by means of a gating voltage \cite{Orus2021}. 
On the other hand, three-dimensional W-C nanostructures can be grown by \ce{Ga+} FIBID as freestanding pick-up loops coupled to a SQUID \cite{Romans2010} and can be designed to support unconventional vortex patterns \cite{Cordoba2019b}.
Recently, K. Lahabi and co-workers have characterized the electronic and magnetic-field dependent properties of JJs created by FEBID \cite{Blom2021}. 
Superconducting properties of planar and 3D superconducting NWs based on \ce{Nb\hyphen C} nanodeposits and fabricated by FEBID and \ce{Ga+} FIBID have also been studied \cite{Porrati2019, Dobrovolskiy2020}, with 3D nanowires exhibiting a higher critical temperature than their 2D counterparts.

The vast knowledge on the properties of \ce{W\hyphen C} and on how to tune them as desired, together with its commercial availability, make the \ce{W(CO)6} precursor the perfect candidate for a broad range of approaches for the fabrication of nanoSQUIDs.
In a single-step process a combination of normal- and superconducting materials can be used to fabricate SQUIDs based on both SNS- and SsS-JJs.

In this work, we present a method to nanofabricate \ce{W\hyphen C} based dc-SQUIDs with two DBs by means of focused \ce{Ga+} ion beam induced deposition on flat \ce{Si}/\ce{SiO2} substrates.
\Cref{sec:experimental} describes the instruments and parameters used to carry out the experiments.
\Cref{ssec:fabrication} outlines the fabrication process that we have developed to fabricate nanoSQUIDs in a single writing step with high reproducibility and yield.
In \cref{ssec:initchar} and \cref{ssec:magchar} the results of the characterization of the electric and magnetic properties of the dc-SQUIDs are outlined.

\section{Experimental}
\label{sec:experimental}

The devices were grown on \ce{Si} substrates covered with a thermally-grown, \SI{300}{\nano\metre} thick \ce{SiO2} surface layer.
Prior to the deposition of the W-C nanoSQUIDs, a supporting Cr/Au structure, comprising the current and voltage leads for the electrical measurements of the devices, was patterned onto the substrates by optical lithography.
A S\"uss MicroTec \emph{MA6} mask aligner, equipped with a \SI{405}{\nano\metre} mercury lamp, has been used to transfer the design to a $\sim\SI{2.8}{\micro\metre}$ thick \ce{MMA} resist layer. 
An electron beam deposition system (BOC EDWARDS \emph{Auto 500}) has been used to metallize the sample with a \SI{5}{\nano\metre} \ce{Cr} and a \SI{50}{\nano\metre} \ce{Au} layer followed by lift-off in acetone.
The fine contacting structure was carried out via EBL in a ThermoFisher Scientific \emph{Helios NanoLab 600} FIB/SEM microscope controlled by a Raith \emph{ELPHY Plus} pattern processor to a \SI{270}{\nano\metre} layer of \ce{PMMA} resist.
The metallization and lift-off steps were then repeated as described above.

The nanofabrication and imaging of the \ce{W\hyphen C} SQUIDs were performed in the same \emph{Helios 600 NanoLab} FIB/SEM microscope, fitted with a \ce{Ga+}ion column and a gas injection system (GIS) for precursor delivery.
The imaging was performed with an electron beam current of \SI{1.4}{\nano\ampere} at an acceleration voltage of \SI{5}{\kilo\volt}.
For the deposition of the \ce{W\hyphen C} material, an ion beam current of \SI{1.5}{\pico\ampere} and an acceleration voltage of \SI{30}{\kilo\volt} were used.
The volume per dose was \SI{8.3e-2}{\micro\metre\cubed\per\nano\coulomb}, the overlap was set to \SI{50}{\percent} and the dwell time was \SI{500}{\micro\second}.
The base pressure of the FIB/SEM chamber was at \SI{e-6}{\milli\bar}, rising to \SI{e-5}{\milli\bar} during the injection of the \ce{W(CO)6} precursor gas.
The nozzle of the GIS was positioned at a vertical distance of \SI{50}{\micro\metre} and a in-plane displacement of \SI{100}{\micro\metre} from the irradiation point.

The low temperature characterization of the magnetotransport properties of the sample was performed in a commercial Quantum Design \emph{Physical Property Measurement System} instrument.
The base temperature for the characterization was \SI{2}{\kelvin}.
The samples were connected to the instrument via ultrasonic wire-bonding of \ce{Al} wires between the \ce{Cr}/\ce{Au} leads and the instrument sample holder.

\section{Results and discussion}
\label{sec:results}

\subsection{Device fabrication}
\label{ssec:fabrication}

The fabrication of the nanodevices has been performed in a series of sequential steps (\cref{fig:Fabrication}), which include the fabrication of both the Josephson junctions and the main body of the nanoSQUID.
In order to obtain the highest possible lateral resolution when depositing the nanowires, the lowest ion beam current of the FIB (\SI{1.5}{\pico\ampere}) was chosen.
The current has been kept at this value for the whole structure.

After the process chamber was flushed with the precursor gas for \SI{20}{\second}, two \SI{50}{\nano\metre}-thick \ce{W\hyphen C} large leads were deposited to carry the injected current from the \ce{Au} leads to the device (\cref{sfig:Fabrication-a}), taking \SI{120}{\second}.
Thereafter, two $\SI{800}{\nano\metre}\times\SI{200}{\nano\metre}$ rectangular pads with a thickness of \SI{50}{\nano\metre} were deposited in contact with the leads fabricated in the previous step, and positioned \SI{300}{\nano\metre} apart from each other (\cref{sfig:Fabrication-b}), taking \SI{60}{\second}.

During the deposition of materials by FIBID, a common problem is the undesired deposition of material in the vicinity of the irradiated area, an issue commonly referred to as \emph{halo}.
In the case of conductive deposits, the halo can carry part of the injected current. This is the reason why the halo deposit must be eliminated to ensure proper device functionality.
We observed a significant amount of halo in between the pads, \emph{i.e.} at the effective loop area of the SQUID.
Thus the fabrication was paused until the precursor gas was completely evacuated from the chamber, taking \SI{30}{\second}, and a short FIB milling step of \SI{2}{\second} of the effective loop area was performed (\cref{sfig:Fabrication-c}) in order to remove the unwanted metallic deposits inside the inner loop area of the SQUID.
Upon gas injection for \SI{20}{\second}, two nanowires acting as DBs were deposited connecting the two pads by their outer edges (\cref{sfig:Fabrication-d}), taking \SI{5}{\second}.
In total, the full fabrication process of the SQUID takes less than \SI{3}{\minute} plus the time required to deposit the leads which strongly depends on their length.
The nanowires have a cross-sectional area of $\SI{50}{\nano\metre}\times\SI{50}{\nano\metre}$. 
The overall nominal loop area of the SQUID is $\SI{300}{\nano\metre}\times\SI{700}{\nano\metre}$.

\subsection{Superconducting properties}
\label{ssec:initchar}

In this section we present the superconducting properties of two identically grown SQUIDs, labelled A and B, in absence of an external magnetic field.
The samples were cooled down to the base temperature of \SI{2}{\kelvin} while constantly injecting a bias current $I_\text{b}=\SI{0.2}{\micro\ampere}$ and measuring the resistance, $R$.
The temperature dependence of the resistance is shown in \cref{sfig:InitialCharacterization-RT}.
Both SQUIDs exhibit a transition to the superconducting regime at $T_\text{c,A}=\SI{4.29}{\kelvin}$ and $T_\text{c,B}=\SI{4.17}{\kelvin}$, respectively.
This is in good agreement with the results found in literature \cite{Sadki2004}.

Thereafter, the current vs. voltage dependence was measured to obtain the critical current, shown in \cref{sfig:InitialCharacterization-IV}.
One can notice several transitions, attributed to, both the contact pads and the DBs.
A quantitative analysis indicates that the critical current of the DBs equals $I_\text{c,A}=\SI{8.51}{\micro\ampere}$ and $I_\text{c,B}=\SI{7.98}{\micro\ampere}$ for sample A and B, respectively. 
Although DB-SQUIDs are expected to exhibit hysteretic behavior due to dissipation of heat in the normal conducting state of the constrictions \cite{Koelle2017} we do not observe a hysteresis in the I-V characteristics.
We attribute the suppression of the hysteresis and the high transition width to an increase of the effective temperature in the noise parameter $\Gamma=k_\text{B}T_\text{eff}/E_\text{J}$ due to noise in the bias current \cite{Clarke2004}.
The normal state resistance of the full structure, \emph{i.e.} above all transitions, is $R_\text{N,A}=\SI{496}{\ohm}$ and $R_\text{N,B}=\SI{493}{\ohm}$ for each sample.
Both the critical current and the critical temperature are very similar for the two nanoSQUIDs, confirming the reproducibility of the fabrication procedure.
However, the transition width of sample B is smaller than that of sample A.

\begin{figure}[!ht]
    \centering
    \includegraphics[width=.5\linewidth]{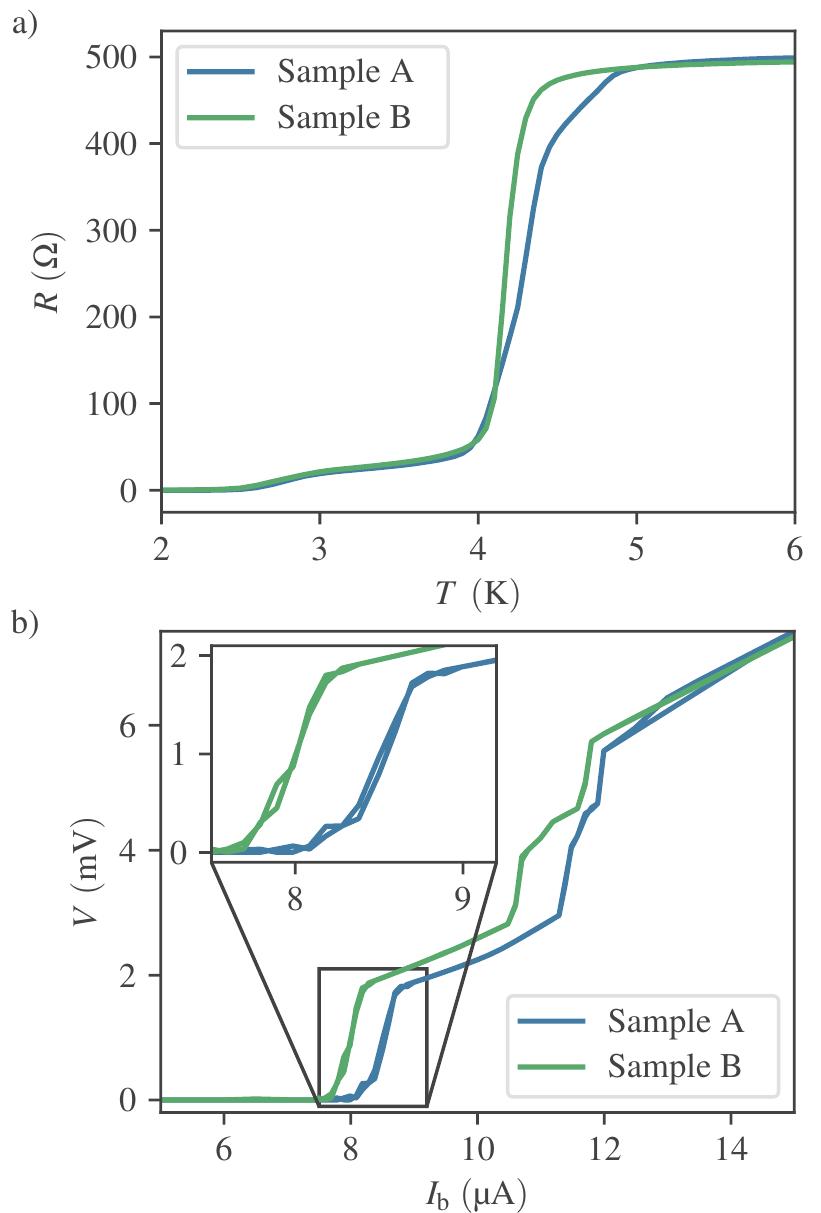}
    \labelphantom{sfig:InitialCharacterization-RT}
    \labelphantom{sfig:InitialCharacterization-IV}
    \caption{%
        a) Temperature dependence of the resistance in the region of the transition from the normal to the superconducting state. %
        b) I-V characteristics of the two samples. %
        The different transitions are attributed to regions of different cross-sectional area.
    }
    \label{fig:InitialCharacterization}
\end{figure}

\subsection{Magnetic response}
\label{ssec:magchar}

\begin{figure*}[ht!]
    \centering
    \includegraphics[width=\linewidth]{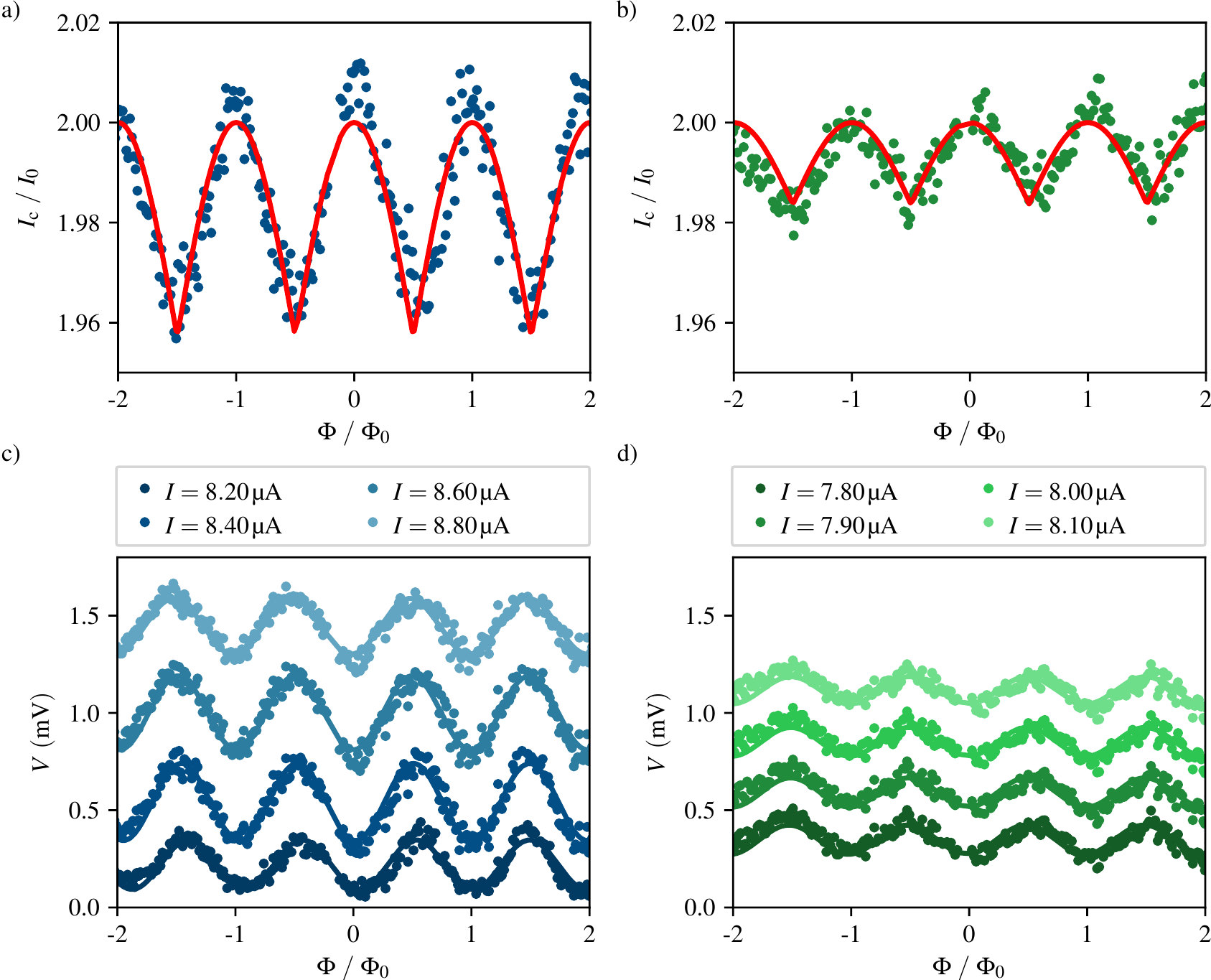}
    \labelphantom{sfig:MagneticCharacterization-IcPhi-A}
    \labelphantom{sfig:MagneticCharacterization-IcPhi-B}
    \labelphantom{sfig:MagneticCharacterization-VPhi-A}
    \labelphantom{sfig:MagneticCharacterization-VPhi-B}
    \caption{
        Electrical response of the devices to an external magnetic field at \SI{2}{\kelvin}. %
        a, b) Critical current in dependence of the magnetic flux threading the SQUID loop for sample A (a) and B (b). %
        $I_\text{c}$ is normalized by the maximal Josephson current of a single junction. 
        The red curve denotes the fit of the data to the model $I_\text{c}=2I_0\vert\cos(\Phi/\Phi_0)\vert$. %
        c, d) Voltage at various bias currents $I\sim I_\text{c}$ in dependence of the magnetic flux threading the SQUID loop for sample A (c) and B (d). %
        In all plots the magnetic flux is normalized by the magnetic flux quantum \si{\phinull}. %
    }
    \label{fig:MagneticCharacterization}
\end{figure*}

In a JJ the macroscopic wavefunctions of the two separated superconductors overlap resulting in a phase-dependent Josephson current flowing through, given by

\begin{equation}
    I_\text{J} = I_0\sin\delta(t)
\end{equation}

\noindent where $I_0$ denotes the maximal Josephson current and $\delta$ the phase difference between the two superconductors.
In a dc-SQUID the flux quantization requires the phase of the two JJs to fulfill the condition

\begin{equation}
    \delta_1 + \delta_2 + 2\pi n = \frac{2\pi}{\Phi_0}\Phi_\text{T}
\end{equation}

\noindent where the total flux is $\Phi_\text{T}=\Phi+LJ$, $L$ being the inductance of the SQUID. 
The loop current $J$ ensures this condition by raising or lowering the total flux by $LJ$.
Assuming a negligible contribution of $LJ$ and $I_{0,1} = I_{0,2} = I_0$ we obtain the dependence of the critical current of the nanostructure on the external flux as

\begin{equation}
    \label{eq:IcofPhi}
    I_\text{c} = 2I_0\left\vert\cos\left(\frac{\pi\Phi}{\Phi_0}\right)\right\vert .
\end{equation}

The quality of a SQUID can be characterized by the screening parameter

\begin{equation}
    \beta_\text{L}=\frac{2LI_0}{\Phi_0}.
\end{equation}

\Cref{eq:IcofPhi} holds for $\beta_\text{L}\ll1$ with a critical current modulation $\Delta I_\text{c}/2I_0=1$.
For a non-negligible influence of $LJ$ the critical current modulation decreases monotonically with an increasing screening parameter $\Delta I_\text{c}/2I_0\approx 1/\beta_\text{L}$.
This property allows one to estimate the inductance $L$ of the SQUID.

To extract $I_\text{c}$ we have measured several I-V characteristics of the SQUIDs at different values of perpendicularly-applied magnetic field.
As the magnetic flux threading the SQUID loop is quantized in integer multiples of the magnetic flux quantum $\Phi_0$, it is possible to attribute each period $\Delta H$ of the oscillation in $I_\text{c}(H)$ to a flux difference of $\Delta \Phi\equiv\Phi_0$.
\Cref{sfig:MagneticCharacterization-IcPhi-A,sfig:MagneticCharacterization-IcPhi-B} respectively show the dependence of the critical current of samples A and B (normalized to $I_0$) on the magnetic flux in units of $\Phi_0$.
The red curve corresponds to the fit of the data to \cref{eq:IcofPhi} augmented by an offset current $I_\text{off}$ in order to account for the non-negligible $\beta_\text{L}$:
\begin{equation}
    \frac{I_\text{c}}{I_0} = \left\vert\cos\left(\frac{\pi\Phi}{\Phi_0}\right)\right\vert+I_\text{off}.
\end{equation}
The period of the oscillation in $I_\text{c}$ is similar in the two samples, with $\Delta H_\text{A}=\SI{64.0}{\oersted}$ and $\Delta H_\text{B}=\SI{56.7}{\oersted}$.
The effective areas $A_\text{eff}=\Phi_0/\mu_0\Delta H$ deducted from this result are $A_\text{eff,A}=\SI{0.323}{\micro\metre\squared}$ and $A_\text{eff,B}=\SI{0.365}{\micro\metre\squared}$. 
These are greater than the geometric loop area of $A_\text{geom}\approx\SI{0.175}{\micro\metre\squared}$ which is to be expected due to the contribution of the kinetic inductance to the total inductance of the SQUID.
Due to the high London penetration depth of \ce{W\hyphen C} and the low thickness of the SQUIDs, the screening of the magnetic field is comparably low, yielding a high kinetic inductance \cite{Koelle2017}.

The amplitude of the oscillations yields the dimensionless screening parameter $\beta_\text{L,A}=\num{47}$ and $\beta_\text{L,B}=\num{122}$, and the inductance of the devices, $L_\text{A}=\SI{11}{\nano\henry}$ and $L_\text{B}=\SI{32}{\nano\henry}$.
The magnetic properties of the two SQUIDs are comparable to one another.
However, the determination of $I_\text{c}$ proved more error-prone due to the steeper transition in the I-V characteristics accounting for the minor differences between the samples.

Commonly, nanoSQUIDs are designed to have a screening parameter $\beta_\text{L}\approx1$ and are furthermore optimized to have a low inductance $L$, which reduces the magnetic flux noise and thus improves the sensitivity to magnetic flux variations \cite{Koelle2017}.
In DB-based SQUIDs based on \ce{Nb} superconducting films ($\lambda_\text{L}=\SI{47}{\nano\metre}$ \cite{Maxfield1965}) inductance values in the \si{\pico\henry} regime are typically achieved \cite{Koelle2017}.

The common operation mode of a dc-SQUID is its use as a flux-to-voltage transducer, where a constant bias current $I_\text{b}\sim I_\text{c}$ is injected and the voltage $V$ is measured, exhibiting a sinusoidal dependence on $\Phi$.
In the vicinity of $\Phi=(1/4\pm n/2)\Phi_0$ a linear dependence of $V$ on $\Phi$ is obtained.
The strongest variation of $V$ for a change of $\Phi$ is characterized by the transfer coefficient 

\begin{equation}
    V_\Phi\vert_{I_\text{b}}=\max\left(\frac{\partial V}{\partial \Phi}\right)_{I_\text{b}}
\end{equation}

Thus, curves of constant $I_\text{b}$ have been extracted from the I-V characteristics.
\Cref{sfig:MagneticCharacterization-VPhi-A,sfig:MagneticCharacterization-VPhi-B} show $V(\Phi)$ curves for various values of $I_\text{b}\sim I_\text{c}$ of sample A and B, respectively.
In sample A the transfer coefficient is $V_\Phi=\SI{1301}{\micro\volt/\Phi_0}$ for a bias current of $I_\text{b}=\SI{8.5}{\micro\ampere}$. 
In sample B we obtain $V_\Phi=\SI{473}{\micro\volt/\Phi_0}$ at $I_\text{b}=\SI{7.9}{\micro\ampere}$.
Typical values of the transfer function are in the range of $\SIrange{10}{100}{\micro\volt/\Phi_0}$ \cite{Koelle2017}.
The high transfer coefficient is attributed to the high normal state resistance of the \ce{W\hyphen C
} nanoSQUIDs around \SI{500}{\ohm}, due to the high resistivity of this material \cite{Sadki2004} in comparison to other structures reported in the literature, which is commonly around $\SIrange{10}{50}{\ohm}$ \cite{Russo2014, Troeman2007}.

\section{Outlook}
\label{sec:outlook}

This fabrication procedure serves as a proof of concept for the fabrication of \ce{W\hyphen C} nanoSQUIDs by means of \ce{Ga+} FIBID.
It has the prospect to be modified and augmented in various ways.
The conduction regime (normal- or superconducting) of the \ce{W\hyphen C} deposit can be controlled by deposition at various substrate temperatures (Cryo-FIBID) \cite{Cordoba2019} or the use of an electron beam (FEBID) at different beam currents \cite{Blom2021}.
Thus the DBs can be readily replaced by a non-superconducting, FIBID-grown metal, resulting in SNS junctions.
Thereby the fabrication of planar instead of sandwich-type SNS-JJ based nanoSQUIDs could be realized.
A metallic heat-sink or a shunt resistor can be also added to the SQUID in a similar manner.

In recent years the development of SQUID on Tip (SOT) probes resulted in a new generation of Scanning SQUID Microscopes (SSMs) with unprecedented resolution and sensitivity for the mapping of the magnetic structure of a given surface \cite{Finkler2010, Anahory2020}.
In this approach a SQUID is positioned on the tip of a pulled quartz tube via a three-step evaporation process. 
Lithographic methods require large, flat substrates and reach their limit with the high aspect ratio of the tip. 
The technique presented here poses a possible alternative approach for the fabrication of an SOT probe on commercially available Atomic Force Microscopy (AFM) cantilevers.
The apex of the tip could be cut with the FIB and thereafter a SQUID could be deposited on the resulting flat area while maintaining the previously discussed flexibility in the SQUID design.

The comparably high inductance can be improved by both, a lower effective loop area and higher film thickness to enhance flux focusing.
With \ce{Ga+} FIBID the feasible linewidth is at around \SI{50}{\nano\metre} and the London penetration depth is \SI{850}{\nano\metre}. 
Recent studies showed that both parameters could be improved by the use of \ce{He+} ions for the deposition of \ce{W\hyphen C} nanowires. 
Nanowires with a linewidth down to \SI{10}{\nano\metre} exhibit a London penetration depth of $\SIrange{400}{812}{\nano\metre}$ \cite{Basset2019, Orus2020b, Cordoba2019c} making the material a promising candidate for the improvement of the process developed in this article.
Further work towards the the optimization of the noise and sensitivity of the \ce{W\hyphen C} SQUIDs is underway. 

\section{Conclusion}
\label{sec:conclusion}

In this work we have successfully fabricated two \ce{W\hyphen C} nanoSQUIDs with an inner loop area of $\SI{300}{\nano\metre}\times\SI{700}{\nano\metre}$ in a fast \ce{Ga+} FIBID-FIB process (<\SI{3}{\minute}).
The SQUIDs exhibit a critical temperature of around \SI{4.2}{\kelvin} and a critical current of around \SI{8}{\micro\ampere} at \SI{2}{\kelvin}.
Albeit the London penetration length of \ce{W\hyphen C} is higher than that of similar devices of other materials, we have clearly observed oscillations of both the critical current and the voltage in dependence of the applied external magnetic field.
The transfer coefficient is comparably high with up to \SI{1301}{\micro\volt\per\phinull}, which we attribute to the high normal state resistance of the devices ($\sim\SI{500}{\ohm}$).

The versatility of FIBID facilitates a high degree of freedom in the geometrical dimensions of the nanostructures and the substrate supporting the nanoSQUID, making the process a promising approach for the fabrication of SOT devices.

\section*{Author Contributions}

F. S. performed the sample growth and the magnetotransport experiments, analyzed the data and wrote the first draft of the manuscript. 
P.O. contributed to the sample growth, data interpretation and the writing of the manuscript. 
S.S. and J.M.D.T. got the funding, supervised the research and contributed to the data interpretation and the writing of the manuscript.

\section*{Conflicts of interest}
There are no conflicts to declare.

\section*{Acknowledgements}

This research was supported by the European Commission under H2020 FET Open grant `FIBsuperProbes' (number 892427), by the Spanish Ministry of Science through the grant PID2020-112914RB-I00, from CSIC through projects PIE202060E187 and Research Platform PTI-001, and by Gobierno de Aragón through the grant E13\_20R with European Social Funds (Construyendo Europa desde Aragón). 
The following networking projects are acknowledged: Spanish Nanolito (RED2018-102627-T) and COST-FIT4NANO (action CA19140). 
Authors would like to acknowledge the use of Servicio General de Apoyo a la Investigación-SAI, Universidad de Zaragoza and the technical support provided by the LMA technicians at Universidad de Zaragoza.
Furthermore we thank Julian Linek of the Physical Institute of the University of Tübingen (EKU) for the fruitful discussions.

\printbibliography
\end{document}